\documentclass[12pt,oneside,reqno]{amsart}
\usepackage[margin=1in]{geometry}		
\usepackage[parfill]{parskip}			
\usepackage{graphicx}
\usepackage{amssymb}
\usepackage{epstopdf}
\usepackage[hidelinks]{hyperref} 
\usepackage[hidelinks]{hyperref} 
\title{Time-Symmetric Resolutions of the Renninger Negative-Result Paradoxes}
\author{Michael B. Heaney\\3182 Stelling Drive\\Palo Alto, CA 94303\\mheaney@alum.mit.edu}
\date{6 September 2023}				
\begin{document}
\maketitle
\begin{abstract}
The 1953 and 1960 Renninger negative-result thought experiments illustrate conceptual paradoxes in the Copenhagen formulation of quantum mechanics. In the 1953 paradox we can infer the presence of a detector in one arm of a Mach-Zehnder interferometer without any particle interacting with the detector. In the 1960 paradox we can infer the collapse of a wavefunction without any change in the state of a detector. I resolve both of these paradoxes by using a time-symmetric formulation of quantum mechanics. I also describe a real experiment that can distinguish between the Copenhagen and time-symmetric formulations.
\end{abstract}
\section{Introduction}
\begin{figure}[htbp]
\begin{center}
\includegraphics[width=8 cm]{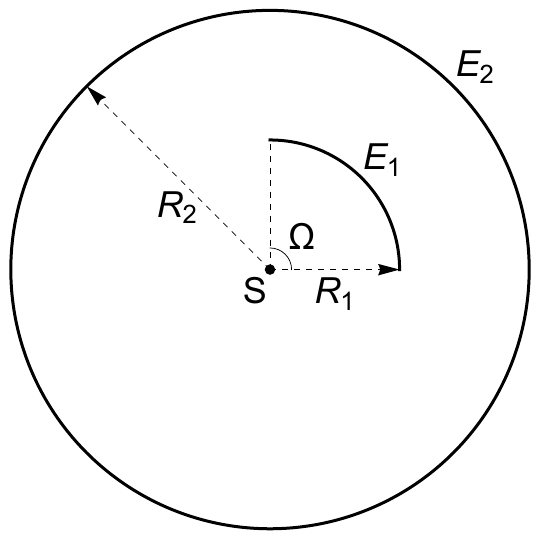}
\caption{The 1960 Renninger negative-result thought experiment. A source S at the center of the sphere $E_2$ emits a single particle whose wavefunction spreads isotropically. The emitted particle is later detected at a localized point either on the inside surface of the sphere sector $E_1$ or on the inside surface of the sphere $E_2$.}
\label{fig1}
\end{center}
\end{figure}

One of the five great problems in theoretical physics is to resolve the conceptual paradoxes in the foundations of the Copenhagen formulation of quantum mechanics, either by making sense of the Copenhagen formulation or developing a different formulation that does make sense~\cite{Smolin}. One of these paradoxes involves negative-result (or interaction-free) measurements~\cite{Dicke}. Such a measurement was proposed by Renninger in 1953 using a Mach-Zehnder interferometer thought experiment~\cite{Renninger1}. In 1960 Renninger proposed a more striking thought experiment using an isotropic point source and a spherical detector~\cite{Renninger2}: see Figure~\ref{fig1}. 
The problem is that in the Copenhagen formulation a wavefunction seems to collapse without interacting with the detectors or leaving any observable trace behind. This is paradoxical.
There is a controversy about what counts as an interaction in these and similar experiments~\cite{Vaidman}. It has also been claimed that what counts as an interaction depends on which formulation of quantum mechanics is being used~\cite{Vaidman,Vaidman2}. In the Copenhagen formulation the absorption of a particle would seem to count as an interaction~\cite{Mitchison}. But what about the collapse to zero of a particle wavefunction upon encountering a detector? What if the spatial wavefunction collapses but the internal state of the particle does not collapse~\cite{Zhou}? I will discuss these issues in the context of the time-symmetric formulation below. 

The structure of the paper is as follows: Section 1 describes the 1960 Renninger negative-result paradox, largely in the words of de Broglie. Section 2 gives a general history and description of time-symmetric formulations of quantum mechanics, and shows how the Copenhagen Born rule is a special case of the time-symmetric transition equation. Section 3 gives a time-symmetric analysis of Renninger's 1960 thought experiment. Section 4 describes the Renninger 1953 negative-result paradox. Section 5 gives a time-symmetric analysis of Renninger's 1953 thought experiment. Section 6 discusses the results and implications of this paper.
\section{The 1960 Paradox}
De Broglie explained the 1960 Renninger negative-result paradox and outlined his proposed resolution as follows~\cite{deBroglie}:
\begin{quote}
In the example which [Renninger] gives, a point source S emits particles isotropically in all directions. A screen $E_1$ in the form of a sector of a sphere centered on S and having a radius $R_1$ is covered on the inside with a substance which indicates the arrival of a particle by a scintillation...Another screen $E_2$, in the form of a complete sphere centered on S and having a radius $R_2 > R_1$, completely surrounds the screen $E_1$. The second sphere is also covered inside with a phosphor. 

Suppose now that the screen $E_1$ subtends a solid angle $\Omega$ at S. The propagation of the wave emitted by S is restricted by the screen $E_1$, and diffraction phenomena occur at the edges of of $E_1$. Notwithstanding the existence of these diffraction phenomena, it is obvious that a particle emitted by the source will have a probability $P_1 = \Omega/4\pi$ of producing a scintillation on $E_1$ and a probability $P_2 = (4\pi-\Omega)/4\pi$ of producing a scintillation on $E_2$. At the instant of emission by the source of a particle with velocity \textit{v}, the emission of the associated wave commences at a time \textit{t=0} and lasts for a finite time $\tau$. The emitted wave $\psi$ forms a spherical shell whose leading edge reaches the screen $E_1$ in a time $t_1=R_1/v$ while the trailing edge reaches the same screen at the time $t_1+\tau$. If, at time  $t_1+\tau$ no scintillation is produced on screen $E_1$, we can be certain that the scintillation will be produced on $E_2$. $P_1$ suddenly becomes zero, and $P_2$ becomes equal to 1. Thus, there will be a sharp change in the amplitude of the wave on the two screens and, according to the usual [Copenhagen] theory, we shall have a special case of the reduction of a probability packet. A particularly paradoxical situation will now exist, since the observer sees nothing at all on screen $E_1$, where nothing has happened. In this experiment the reduction of the probability packet is quite incomprehensible. It is, in fact, impossible to accept that this reduction is due to the increase of knowledge of the observer who has observed nothing, nor to a device---here screen $E_1$---which registered nothing. 

The situation becomes clearer if we accept that the source emits a particle which remains closely associated with a wave, but which has a definite position at each time and, consequently, a definite trajectory. The trajectory should be closely linked with the propagation of the wave and should be influenced by it. It can be accepted that, at least on average, these trajectories are straight lines starting from S, except for the immediate vicinity of the edges of screen $E_1$, which constitute an obstacle to propagation of the wave and give rise to diffraction, thus producing local modifications of the trajectories. On the whole, it can be said that the number of possible trajectories emanating from S and terminating on $E_1$ is proportional to $\Omega$ whilst the number of trajectories emanating from S and reaching $E_2$, whether after a rectilinear trajectory or a trajectory which has been disturbed by diffraction at the edges of screen $E_1$, is proportional to $4\pi-\Omega$. We thus find the probabilities $P_1 = \Omega/4\pi$ and $P_2 = (4\pi-\Omega)/4\pi$ of the arrival of the particle at either $E_1$ or $E_2$. If no scintillation is produced on $E_1$ in the time $t_1+\tau$, which is the time taken by the [trailing edge of the] wave to reach the sphere of radius $R_1$, then we may be sure that the trajectory followed by the particle is not one of those terminating on $E_1$. There would thus be a sudden change to $P_1=0$ and $P_2=1$. This sudden change will represent simply a change in our knowledge of the trajectory of the particle. This removes the incomprehensible effect of the mind of the observer on the particle since there is no scintillation on $E_1$. As far as the presence of "measuring devices" is concerned, the screen $E_1$ is simply an obstacle to the propagation of the wave and thus influences the \textit{possible} trajectories by stopping certain trajectories and giving rise to diffraction. This interpretation is very clear and much more comprehensible than the one based upon a mysterious effect which imposes on the particle the simple \textit{possibility} that it might become localised on $E_1$. How can we possibly imagine that a possibility which has never become real would have such an effect?
\end{quote}
De Broglie then goes on to describe the details of his alternative formulation of quantum mechanics, now known as the pilot wave theory. I will instead explain and resolve Renninger's thought experiment using an alternative time-symmetric formulation of quantum mechanics.
\section{Time-Symmetric Formulations of Quantum Mechanics}
Time-symmetric explanations of quantum behavior predate the discovery of the Schr\"{o}dinger equation~\cite{Tetrode} and have been developed many times over the past century~\cite{Friederich}. The time-symmetric formulation used in this paper has been described in detail and compared to other time-symmetric theories before~\cite{HeaneyA,HeaneyB,HeaneyC,HeaneyD}. The key ideas are that \textit{transitions} between specified quantum states are the basic objects of interest in quantum mechanics, these transitions are fully described by transition amplitude densities, and wavefunction collapse never occurs. These ideas were originally developed by Feynman~\cite{Feynman,QED}. In addition, I postulate that particle sources spontaneously emit isotropic retarded waves, particle detectors spontaneously emit isotropic advanced waves, and a transition only occurs when these two types of waves overlap at a source and a detector. These ideas are very similar to ideas that were originally developed by Cramer~\cite{Cramer,Cramer2}.

The time-symmetric theory used in this paper postulates that the transition of a single free particle is described by the algebraic product of a retarded wavefunction $\psi(\vec{r},t)$ which satisfies the initial conditions and evolves in time according to the retarded Schr\"{o}dinger equation
\begin{equation}
i\frac{\partial\psi}{\partial t}=-\frac{1}{2}\nabla^2\psi,
\label{eq.1}
\end{equation}
and an advanced wavefunction $\phi^\ast(\vec{r},t)$ which satisfies the final conditions and evolves in time according to the advanced Schr\"{o}dinger equation
\begin{equation}
-i\frac{\partial\phi^\ast}{\partial t}=-\frac{1}{2}\nabla^2\phi^\ast,
\label{eq.2}
\end{equation}
where we assume the particle mass $m=1$ and use natural units where $\hbar=1$. These two equations are the low energy limits of the relativistic Klein-Gordon equation~\cite{HeaneyB}. 

The equation of continuity can be obtained as follows. If we multiply Equation~\ref{eq.1} on the left by $\phi^\ast$, multiply Equation~\ref{eq.2} on the left by $\psi$, take the difference of the two resulting equations, and rearrange terms, we get
\begin{equation}
\frac{\partial}{\partial t}(\phi^\ast\psi)+\nabla\cdot\left[ \frac{1}{2i}\left(\phi^\ast\nabla\psi-\psi\nabla\phi^\ast \right)\right]=0.
\label{eq.3}
\end{equation}
Now we will define $\rho_s(\vec{r},t)$ as
\begin{equation}
\rho_s\equiv \phi^\ast\psi,
\label{eq.4}
\end{equation}
and define $\vec{j_s}(\vec{r},t)$ as
\begin{equation}
\vec{j}_s\equiv\frac{1}{2i}\left(\phi^\ast\nabla\psi-\psi\nabla\phi^\ast\right),
\label{eq.5}
\end{equation}
to get a local conservation law
\begin{equation}
\frac{\partial\rho_s}{\partial t}+\nabla\cdot\vec{j}_s=0.
\label{eq.6}
\end{equation}
Note that both $\rho_s(\vec{r},t)$ and $\vec{j_s}(\vec{r},t)$ are generally complex functions, and therefore cannot be interpreted as a real probability density and a real probability density current. Instead, we will interpret $\rho_s(\vec{r},t)$ as the amplitude density for a transition defined as an isolated, individual physical system that starts with maximally specified initial conditions, evolves in space-time, then ends with maximally specified final conditions. The complex function $\rho_s(\vec{r},t)$ is called the transition amplitude density. This complex function is also known as the transition amplitude density in the Copenhagen formulation. We will also interpret $\vec{j_s}(\vec{r},t)$ as the transition amplitude current density. 

Integrating Equation~\ref{eq.6} over all space gives
\begin{equation}
\iiint_{-\infty}^{+\infty}\frac{\partial\rho_s}{\partial t}dV+\iiint_{-\infty}^{+\infty}\nabla\cdot\vec{j}_sdV=0.
\label{eq.7}
\end{equation}
We can use Gauss's theorem to express the second term as
\begin{equation}
\iint_{A}\vec{j}_s\cdot d\vec{A}
\label{eq.8}
\end{equation}
where $A$ is a closed surface at $\vec{r}=\pm\infty$. If we assume the wavefunctions $\psi(\vec{r},t)$ and $\phi^\ast(\vec{r},t)$ are normalized and go to 0 at $\vec{r}=\pm\infty$ this term goes to 0, leaving
\begin{equation}
\frac{d}{dt}\iiint_{-\infty}^{+\infty}\rho_s(\vec{r},t)dV=0
\label{eq.9}
\end{equation}
where we have moved the derivative outside of the integral. We will now define
\begin{equation}
A_s \equiv \iiint_{-\infty}^{+\infty}\rho_s(\vec{r},t)dV,
\label{eq.10}
\end{equation}
where by Equation~\ref{eq.9} $A_s$ is a constant, independent of time.

The time-symmetric formulation interprets $A_s$ as the amplitude for an isolated, individual physical system to start with maximally specified initial conditions, evolve in space-time, then end with maximally specified final conditions. The complex number $A_s$ is called the transition amplitude. This predicts that the volume under the curve of the real part of $\rho_s(x,t)$ is conserved, and the volume under the curve of the imaginary part of $\rho_s(x,t)$ is also conserved. 
The transition probability $P_s$ can then be defined by
\begin{equation}
P_s\equiv A_s^\ast A_s
\label{eq.11}
\end{equation}
where $P_s$ is also a constant, independent of time. The time-symmetric formulation interprets $P_s$ as the probability that an isolated, individual physical system will start with a given set of maximally specified initial conditions, evolve in space-time, then end with a given set of maximally specified final conditions. This is the conditional probability that a particle with the given initial conditions will later be found with the given final conditions. Since $P_s$ is time-independent, it is also the probability that a particle with the given final conditions would have been found earlier with the given initial conditions. This time symmetry is generally true for quantum transition probabilities. 

Let us consider some examples of the transition amplitude $A_s$ and the transition probability $P_s$. First, consider a one-dimensional infinite square well of length $a$. The stationary states are
\begin{equation}
\xi_n(x,t)=\sqrt{\frac{2}{a}}\sin\left(\frac{n\pi}{a}x\right)\exp\left[-i\frac{n^2\pi^2}{2a^2}t\right],
\label{eq.12}
\end{equation}
where $n=1,2,3...$, we assume the particle mass $m=1$, and use natural units where $\hbar=1$. Choose $\psi(x,t)=\xi_1(x,t)$ and $\phi^\ast(x,t)=\xi^\ast_1(x,t)$ so 
\begin{equation}
\rho_s=\phi^\ast\psi=\frac{2}{a}\sin^2\left(\frac{\pi}{a}x\right).
\label{eq.13}
\end{equation}
Then integrating $\rho_s(x)$ over the square well gives 
\begin{equation}
A_s = \int_{0}^{a}\rho_s(x)dx =1
\label{eq.14}
\end{equation}
and $P_s=A^\ast_sA_s=1$ as expected, since the initial and final states perfectly overlap. As a second example, consider choosing $\psi(x,t)=\xi_1(x,t)$ and $\phi^\ast(x,t)=\xi^\ast_2(x,t)$ so
\begin{equation}
\rho_s=\frac{2}{a}\sin\left(\frac{\pi}{a}x\right)\sin\left(\frac{2\pi}{a}x\right)\exp\left[i\frac{3\pi^2}{2a^2}t\right].
\label{eq.15}
\end{equation}
Then integrating $\rho_s(x,t)$ over over the square well gives 
\begin{equation}
A_s = \int_{0}^{a}\rho_s(x,t)dx =0
\label{eq.16}
\end{equation}
and $P_s=A^\ast_sA_s=0$ as expected, since different stationary states are orthogonal to each other. A third example, involving two different stationary but time dependent gaussians, is given in Section 4. In that case, the calculated transition probability is $P_s=3\times 10^{-10}$. We see that the transition probabilities can range from 0 to 1 and are time-independent.

One of the postulates of the Copenhagen formulation is the Born rule
\begin{equation}
\rho=\psi^\ast(\vec{r},t)\psi(\vec{r},t),
\label{eq.17}
\end{equation}
where $\rho(\vec{r},t)$ is the probability density for finding the particle at $\vec{r}$ at time $t$. This is a special case of the transition amplitude density when the initial state is $\psi(\vec{r},t)$ and the final state $\phi^\ast(\vec{r},t)$ is a delta function $\delta(\vec{r}\medspace',t')$:
\begin{equation}
\rho_s=\delta(\vec{r}\medspace',t')\psi(\vec{r},t)=\psi(\vec{r}\medspace',t'),
\label{eq.18}
\end{equation}
\begin{equation}
\rho^\ast_s=\delta(\vec{r}\medspace',t')\psi^\ast(\vec{r},t)=\psi^\ast(\vec{r}\medspace',t'),
\label{eq.19}
\end{equation}
and then redefining variables ($\vec{r}\medspace'\rightarrow \vec{r}$, $t'\rightarrow t$) gives
\begin{equation}
\rho=\rho^\ast_s\rho_s=\psi^\ast(\vec{r},t)\psi(\vec{r},t)
\label{eq.20}
\end{equation}
which is the Copenhagen Born rule. Note that although the amplitude density of Equation~\ref{eq.17} is time dependent, the integral over all space of Equation~\ref{eq.17} is
\begin{equation}
\iiint_{-\infty}^{+\infty}\psi^\ast(\vec{r},t)\psi(\vec{r},t)dV=1
\label{eq.21}
\end{equation}
which is the probability of finding the particle somewhere in space. This shows that the Copenhagen formulation Born rule is consistent with the time symmetric formulation: the amplitude densities depend on time and space, but when integrated over all space the results are independent of time.

Note that $\phi^\ast\psi$ is a time-symmetric formulation amplitude density, giving the amplitude for a transition from state $\psi$ to state $\phi$. It is not the same as the Born rule probability amplitude density $\psi$ of finding the particle to be at some location in space. For example, the Born rule probability density $\psi^\ast\psi$ must always be equal to one when integrated over all space. But the transition amplitude density $\phi^\ast\psi$ will usually not be equal to one when integrated over all space, because $\phi^\ast$ and $\psi$ need not perfectly overlap.

The time-symmetric formulation tries to treat $\psi$ and $\phi^\ast$ on an equal footing. In practice in the laboratory, we can usually manipulate $\psi$ because it originates in our present. It is sometimes possible to manipulate $\phi^\ast$ in the laboratory. For example, consider the Mach-Zehnder Interferometer of Figure 3. By changing the phase in one of the arms, we can change the future final state $\phi^\ast$ from a particle always detected in detector $D_1$ to a particle always detected in detector $D_2$. But in general, we usually cannot manipulate $\phi^\ast$ because our world is asymmetric in time and $\phi^\ast$ originates in our future.
\section{The Time-Symmetric Explanation of Renninger's 1960 Thought Experiment}
For easier visualization we will assume the experiment shown in Figure~\ref{fig1} is two-dimensional and both wavefunctions are stationary Gaussians with initial and final standard deviations $\sigma=1$. The retarded stationary Gaussian is\\
\begin{equation}
\psi(x,y,t)\equiv\left(\frac{2}{\pi}\right)^{1/2}\left(\frac{1}{i(t-t_i)+2}\right)exp\left[-\frac{(x-x_i)^2+(y-y_i)^2}{2i(t-t_i)+4}\right],
\label{eq.22}
\end{equation}
where $(x,y)$ is the location of the particle, $t$ is the time, $(x_i,y_i,t_i)=(0,0,0)$ is the emission location and time, all masses are set to $1$, and natural units are used: $\hbar=1$. 

The advanced stationary Gaussian is\\
\begin{equation}
\phi^\ast(x,y,t)\equiv\left(\frac{2}{\pi}\right)^{1/2}\left(\frac{1}{i(t_f-t)+2}\right)exp\left[-\frac{(x_f-x)^2+(y_f-y)^2}{2i(t_f-t)+4}\right],
\label{eq.23}
\end{equation}
where $(x,y)$ is the location of the same particle, $t$ is the time, $(x_f,y_f,t_f)=(0,-60,28)$ is the detection location and time, all masses are set to $1$, and natural units are used: $\hbar=1$.

\begin{figure}[htbp]
\begin{center}
\includegraphics[width=13 cm]{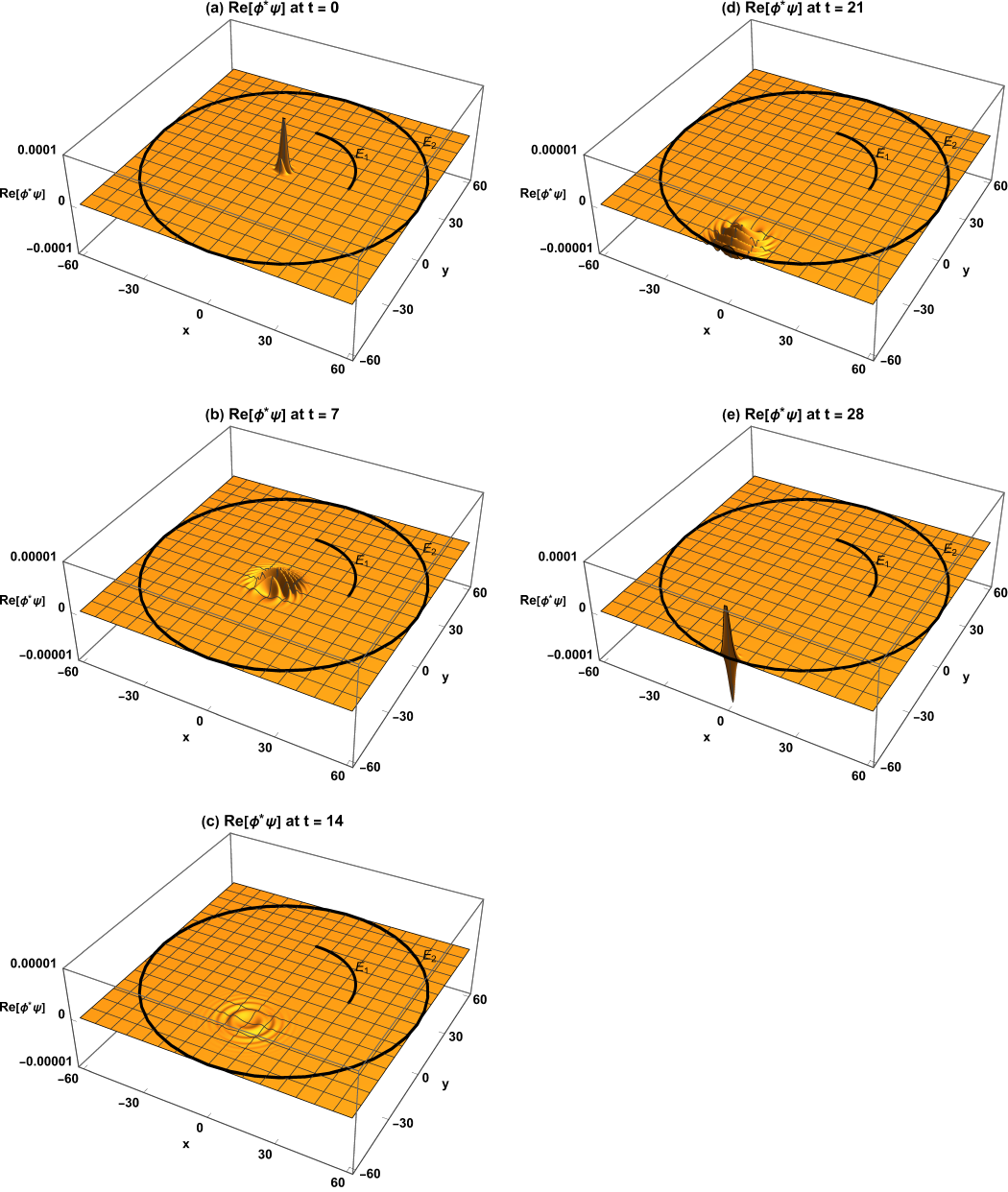}
\caption{The time-symmetric formulation explanation of Renninger's 1960 negative-result experiment in two dimensions, with a single particle emitted from (0,0) and detected on the circle $E_2$ at (0,-60). (\textbf{a}) The transition amplitude density $\phi^\ast\psi$ is localized at the source S. (\textbf{b,c,d}) $\phi^\ast\psi$ has left S and is traveling towards the circle $E_2$. Note that the scale on the vertical axis varies between graphs. (\textbf{e}) $\phi^\ast\psi$ arrives at the circle $E_2$ as a localized transition amplitude density and produces a scintillation. The transition amplitude density diverges from the source and converges to the detector in a time-symmetric manner, without hitting the arc $E_1$. If the detector had been located in the upper right quadrant of the circle $E_2$, the transition amplitude density would have diffracted around the arc $E_1$.}
\label{fig2}
\end{center}
\end{figure}

Figure~\ref{fig2} shows how the transition amplitude density $\phi^\ast\psi$ evolves over time, assuming the initial condition is localization at source S at the origin and the final condition is localization at the outer circle $E_2$ at $(x,y)=(0,-60)$. There is no wavefunction collapse between the initial and final conditions. 

The time-symmetric formulation assumes the probability $P_s$ for this transition is $P_s=A_s^\ast A_s$, where the subscript $s$ denotes the time-symmetric theory and the amplitude $A_s$ for the transition is
\begin{equation}
A_s=\int_{-\infty}^{\infty}\phi^\ast(x,y,t)\psi(x,y,t)dxdy,
\label{eq.24}
\end{equation}
where $t$ is a variable. Plugging in numbers gives a time-symmetric transition probability $P_s=3\times 10^{-10}$. The Copenhagen formulation assumes $t=t_f$ and gets the same numerical result for the transition probability.
\section{The 1953 Paradox}
\begin{figure}[htbp]
\begin{center}
\includegraphics[width=8 cm]{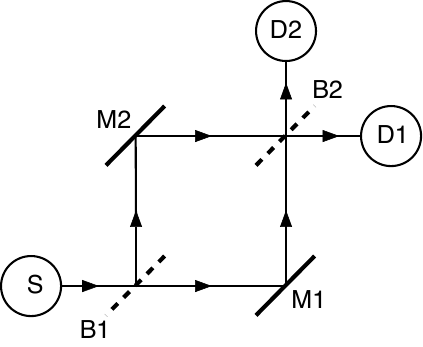}
\caption{The 1953 Renninger negative-result thought experiment. A Mach-Zehnder Interferometer (MZI) is formed by a single-particle source S, two single-particle detectors $D1$ and $D2$, two beam-splitters $B1$ and $B2$, and two mirrors $M1$ and $M2$. The source S emits a single particle whose wavefunction is a traveling gaussian. The MZI is constructed such that when all arms of the interferometer are open the emitted particle always goes to detector $D1$.}
\label{fig3}
\end{center}
\end{figure}

In 1953 Renninger proposed a negative-result thought experiment using a Mach-Zehnder Interferometer [MZI]~\cite{Renninger1}: see Figure~\ref{fig3}. The MZI is constructed such that when both arms of the interferometer are open the emitted particle always goes to detector $D1$. In the Copenhagen formulation, this implies the particle's wavefunction must have taken both arms of the MZI. In the time-symmetric formulation, this implies the particle's transition amplitude density must have taken both arms of the MZI. But if a third detector $D3$ is surreptitiously placed between beam-splitter $B1$ and mirror $M2$ (see Figure~\ref{fig4}) there will be three possible outcomes: the particle is detected in detector $D1$ with probability $1/4$, the particle is detected in detector $D2$ with probability $1/4$, or the particle is detected in detector $D3$ with probability $1/2$. In the cases where the particle is detected in detector $D2$, we can infer the presence of detector $D3$ without any particle interaction with detector $D3$, which is a paradox. Alternatively, if we know that detector $D3$ is blocking the upper arm of the MZI, and we wait until detector $D3$ could have detected the particle but see no detection, then we can conclude that the particle's wavefunction is only taking the lower arm of the MZI. As in the 1960 thought experiment, we have localized the particle's wavefunction without any interaction with the particle, which is a paradox. A similar thought experiment was later described by Elitzur and Vaidman~\cite{Elitzur}.
\section{The Time-Symmetric Explanation of Renninger's 1953 Thought Experiment}
\begin{figure}[htbp]
\begin{center}
\includegraphics[width=13 cm]{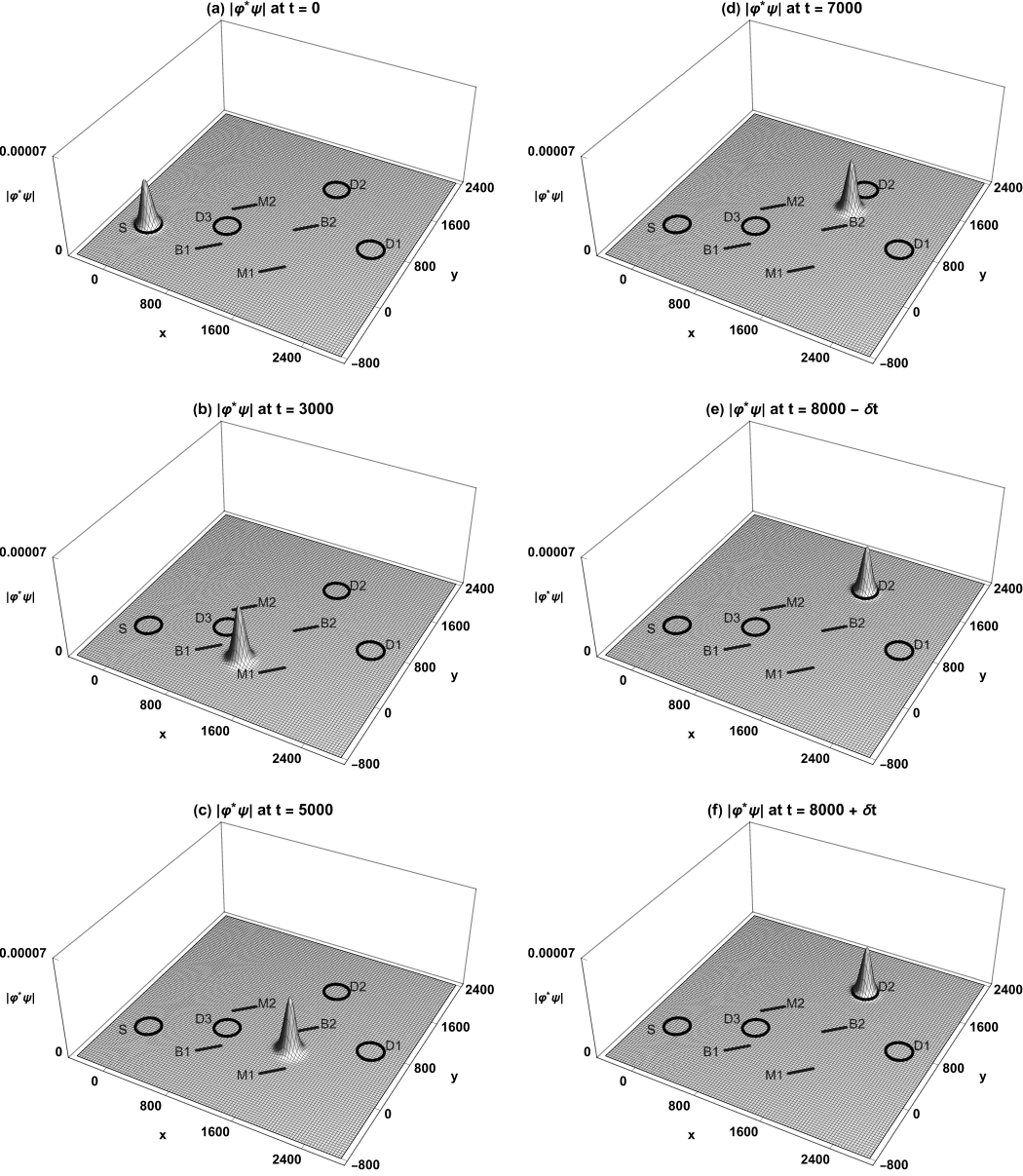}
\caption{The time-symmetric formulation explanation of Renninger's 1953 negative-result experiment, with a single particle emitted from source S and detected at detector $D2$. (\textbf{a}) The transition amplitude density $\phi^\ast\psi$ is emitted from the source S. (\textbf{b}) The transition amplitude density $\phi^\ast\psi$ has passed through the beam-splitter $B1$. Note that transition amplitude densities are not necessarily split by beam-splitters. (\textbf{c}) $\phi^\ast\psi$ has reflected from mirror $1$ and is traveling towards beam-splitter $B2$. (\textbf{d}) $\phi^\ast\psi$ has passed through beam-splitter $B2$ and is traveling towards detector $D2$. (\textbf{e}) $\phi^\ast\psi$ has reached detector $D2$ but has not yet been detected. (\textbf{f}) $\phi^\ast\psi$ has been detected by detector $D2$. Note that wavefunction collapse does not occur.}
\label{fig4}
\end{center}
\end{figure}

The time-symmetric formulation postulates that particle sources spontaneously emit isotropic retarded waves, particle detectors spontaneously emit isotropic advanced waves, and a transition only occurs when these two types of waves overlap at a source and a detector. The presence of detector $D3$ between beam-splitter $B1$ and mirror $M2$ prevents the retarded wave from source S from overlapping with the advanced waves from detectors $D1$ or $D2$, so no transition amplitude density can form along the upper arm of the MZI. But transition amplitude densities can still form between source S and detector $D1$ or detector $D2$ along the lower arm of the MZI. Figure~\ref{fig4} shows an example of a transition amplitude density moving between source S and detector $D2$ along the lower arm of the MZI. 

The retarded traveling gaussian is
\begin{equation}
\psi(x,y,t)=\frac{50 \sqrt{\frac{2}{\pi }} \exp \left[0.4 i (-0.2 (t-t_i)+x-x_i )-\frac{(-0.4 (t-t_i)+x-x_i)^2+(y-y_i)^2}{10000+2 i (t-t_i)}\right]}{5000+i (t-t_i)}
\label{eq.25}
\end{equation}
where $t_i$ is the initial time and $x_i$ and $y_i$ are the initial position.\\
The advanced traveling gaussian is
\begin{equation}
\phi^\ast(x,y,t)=\frac{50 \sqrt{\frac{2}{\pi }} \exp \left[-0.4 i (-0.2 (t_f-t)-x+x_f)-\frac{(-0.4 (t_f-t)-x+x_f)^2+(y_f-y)^2}{10000-2 i (t_f-t)}\right]}{5000-i (t_f-t)}.
\label{eq.26}
\end{equation}
where $t_f$ is the final time and $x_f$ and $y_f$ are the final position.
\section{Discussion} 
The time-symmetric formulation resolves the 1960 Renninger negative-result paradox because both the initial and final states must be specified to apply the theory, the transition amplitude density does not collapse, and the transition amplitude density travels as a localized beam between the initial and final states, terminating on either the inside surface of the sphere sector $E_1$ or the inside surface of the sphere $E_2$. For repeated experiments, we can estimate the probabilities to be $P_1 = \Omega/4\pi$ and $P_2 = (4\pi-\Omega)/4\pi$. If there is no particle detection by the time $t_1+\tau$, the probabilities suddenly change to $P_1=0$ and $P_2=1$. But there is no associated change in the transition amplitude density. This sudden change in probabilities simply reflects our change in knowledge of the trajectory of the transition amplitude density. 

The time-symmetric formulation has the additional benefit of being consistent with the classical limit of Renninger's 1960 thought experiment. As the quantum particle becomes more massive, with a shorter de Broglie wavelength, and starts behaving more like a classical particle, it will always go to either the inner sphere section or the outer sphere in a straight trajectory with a narrow dispersion. There is a logical continuity between its behavior in the quantum and classical regimes, in contrast to the Copenhagen formulation predictions.

The time-symmetric formulation resolves the 1953 Renninger negative-result paradox because a retarded wave from a particle source and an advanced wave from a particle detector must overlap at a source and detector for a transition amplitude density to form. Neither a retarded wave by itself nor an advanced wave by itself will trigger a detector or cause a particle transition. Since the upper arm of the MZI between the source S and the detectors $D1$ and $D2$ is blocked, a transition amplitude density cannot form between the source S and the detectors $D1$ or $D2$ in that arm. But a transition amplitude density can still form in the lower arm of the MZI between the source S and the detectors $D1$ or $D2$. The retarded and advanced waves essentially tell the particle which pathways are blocked and open before the particle takes the pathways.

Note that in both thought experiments, in the time-symmetric formulation, the source emits an isotropic retarded wave and the detector emits an isotropic advanced wave that can hit the detectors and sources. These may count as interactions. But the retarded and advanced waves each by themselves cannot trigger a detector. In contrast, in the Copenhagen formulation, when the retarded wave by itself hits the detectors, it can collapse to a particle. The Copenhagen formulation does not explain why it chooses one detector and not a different detector, while the time-symmetric formulation does.

The time-symmetric formulation also resolves the more general Copenhagen formulation paradox of a nonlocalized wavefunction instantaneously collapsing into a localized wavefunction at a detector. In order to conserve momentum this collapse must be instantaneous in all reference frames, in clear conflict with the special theory of relativity. In the time-symmetric formulation the transition amplitude density is localized at the source, partly delocalizes as it approaches the halfway point between source and detector, then relocalizes as it continues to the detector. No wavefunction collapse is required.

One might wonder if a theory based on transition amplitude densities will be able to reproduce all of the predictions of the Copenhagen formulation. In 1932 Dirac showed that all the experimental predictions of the Copenhagen formulation of quantum mechanics can be formulated in terms of transition probabilities~\cite{Dirac}. The time-symmetric formulation inverts this fact by postulating that quantum mechanics is a theory which experimentally predicts \textit{only} transition probabilities. This implies the time-symmetric formulation has the same predictive power as the Copenhagen formulation.

The Copenhagen formulation has several asymmetries in time: only the initial conditions of the wavefunction are specified, the wavefunction is evolved only forward in time, the transition probability is calculated only at the time of measurement, wavefunction collapse happens only at the time of measurement, and wavefunction collapse happens only forwards in time. This seems unphysical: shouldn't the fundamental laws of nature be time-symmetric? Consider the details of a specific example: according to the Copenhagen formulation, Equation~\ref{eq.24} must be evaluated only at the time of the collapse. In contrast, according to the time-symmetric formulation, the transition amplitude of Equation~\ref{eq.24} can be evaluated at any time. But the two transition amplitudes give the same results. The fact that the transition amplitude need not be evaluated at a special time shows that quantum mechanics has more intrinsic symmetry than allowed by the Copenhagen formulation. Heisenberg said "Since the symmetry properties always constitute the most essential features of a theory, it is difficult to see what would be gained by omitting them in the corresponding language~\cite{Heisenberg}.'' The intrinsic time symmetry of a quantum transition is built into the time-symmetric formulation, but is not present in the Copenhagen formulation.

The Copenhagen formulation predicts a rapid oscillating motion of a free particle in empty space. Schr\"odinger discovered the theoretical possibility of this rapid oscillating motion in 1930, naming it zitterbewegung~\cite{Schroedinger}. This prediction of the Copenhagen formulation is inconsistent with Newton's first law, since it implies a free particle does not move with a constant velocity. The time-symmetric formulation predicts zitterbewegung will never occur~\cite{HeaneyA}. Direct measurements of zitterbewegung are beyond the capability of current technology, but future technological developments should allow measurements to confirm or deny its existence, thereby distinguishing between the Copenhagen formulation and the time-symmetric formulation. One possible future experiment to directly measure zitterbewegung would be to trap an electron in a Penning trap in a gaussian ground state, then remove the trap fields and use antennae to search for zitterbewegung radiation.

The Copenhagen formulation assumes an isolated, individual physical system is maximally described by a retarded wavefunction and maximally specified initial conditions. The time-symmetric formulation assumes a complete experiment is maximally described by the time-symmetric amplitude density $\rho_s(\vec{r},t)$, which is composed of a retarded wavefunction and an advanced wavefunction, and maximally specified initial and final conditions. The existence of both retarded and advanced wavefunctions in the time-symmetric formulation does not imply that particles can travel at superluminal speeds. In the time-symmetric formulation every particle is represented by algebraic products of an advanced wavefunction and a retarded wavefunction, so the particle cannot travel to space-time locations that the retarded wavefunction cannot reach, and relativistic wave equations limit the velocity of the retarded wavefunction to less than $c$. Conversely, the particle cannot travel to space-time locations that the advanced wavefunction cannot reach. This is a type of symmetrical forward and backward causality: what happens during an experiment depends on what happened at the start of the experiment, and what will happen at the end of the experiment. This suggests the past, present, and future have equal status. This is implicit in the time-symmetric postulates, and consistent with the block universe view and with the special theory of relativity. 

The Copenhagen formulation postulate that an individual particle is maximally described by a retarded wavefunction and maximally specified initial conditions means the Copenhagen formulation is a "presentist" theory, where only the present moment is real: the past is no longer real, and the future is not yet real. A "presentist" theory is equivalent to a three-dimensional world, which changes as time passes. The time symmetric formulation postulates that a complete experiment is maximally described by the time symmetric amplitude density $\phi^\ast\psi$, which is composed of a retarded wavefunction and an advanced wavefunction, and incorporates maximally specified initial and final conditions. This means the time symmetric formulation is an "eternalist" theory, where the past, present, and future are equally real. The "eternalist" theory is equivalent to a four-dimensional world, where time is just another parameter, like position. It is an experimental fact, proven by many experiments confirming the special theory of relativity, that the world is four-dimensional, not three-dimensional.

Finally, the time-symmetric formulation may be able to resolve other negative-result or interaction-free paradoxes such as counterfactual quantum computation. A future paper will address these topics.
 
\end{document}